\documentclass[12pt]{article}
\usepackage{amssymb,amsmath,cite}
\catcode`\@=11

\@addtoreset{equation}{section}
\def\theequation{\arabic{section}.\arabic{equation}}
     
\catcode`\@=11
\def\thesection{\arabic{section}}
\def\appendix{\setcounter{section}{0}
        \def\thesection{Appendix \Alph{section}}
        \def\theequation{\Alph{section}.\arabic{equation}}}
\def\section{\@startsection{section}{1}{\z@}{3.5ex plus 1ex minus
   .2ex}{2.3ex plus .2ex}{\large\bf}}

\long\def\@makefntext#1{\parindent 0cm\noindent
\hbox to 1em{\hss$^{\@thefnmark}$}#1}

\def\rref#1{(\ref{#1})}
\newcommand{\beq}{\begin{equation}}
\newcommand{\eeq}{\end{equation}}

\begin{document}
\begin{titlepage}
\begin{flushright}
UCD-2003-10\\
September 2003\\
revised November 2003\\
gr-qc/0310002\\
\end{flushright}
\vspace{.1in}
\begin{center}
{\Large\bf
 Peaks in the Hartle-Hawking Wave Function\\[.6ex]
 from Sums over Topologies}\\[3ex]

M.~A{\sc nderson}\footnote{\it email: anderson@math.sunysb.edu}\\
       {\small\it Department of Mathematics, SUNY at Stony Brook}\\
       {\small\it Stony Brook, NY 11794, USA}\\[2ex] 
S.~C{\sc arlip}\footnote{\it email: carlip@dirac.ucdavis.edu}\\
       {\small\it Department of Physics, University of California }\\
       {\small\it Davis, CA 95616, USA}\\[2ex] 
J.~G.~R{\sc atcliffe}\footnote{\it email: John.G.Ratcliffe@Vanderbilt.edu}\\
       {\small\it Department of Mathematics,Vanderbilt University }\\
       {\small\it Nashville, TN 37240, USA}\\[2ex] 
S.~S{\sc urya}\footnote{\it email: ssurya@rri.res.in }\\
      {\small\it Raman Research Institute, C.V.\ Raman Avenue, Sadashivanagar}\\
      {\small\it Bangalore 560 080, INDIA}\\[2ex] 
S.~T.~T{\sc schantz}\footnote{\it email: Steven.T.Tschantz@Vanderbilt.edu}\\ 
       {\small\it Department of Mathematics,Vanderbilt University }\\
       {\small\it Nashville, TN 37240, USA} 
\end{center}
\vspace*{-1ex}
\begin{center}
{\large\bf Abstract}
\end{center}
\begin{center}
\begin{minipage}{4.75in}
{\small
Recent developments in ``Einstein Dehn filling'' allow the 
construction of infinitely many Einstein manifolds that 
have different topologies but are geometrically close to 
each other.  Using these results, we show that for many
spatial topologies, the Hartle-Hawking wave function for 
a spacetime with a negative cosmological constant develops 
sharp peaks at certain calculable geometries.  The peaks 
we find are all centered on spatial metrics of constant 
negative curvature, suggesting a new mechanism for 
obtaining local homogeneity in quantum cosmology.}
\end{minipage}
\end{center}
\end{titlepage}
\addtocounter{footnote}{-5}

\section{Introduction \label{intro}} 

Quantum cosmology is a difficult subject, not least because we do 
not yet have a complete quantum theory of gravity.  In the absence 
of such a theory, cosmologists must rely on plausible, but necessarily 
speculative, approaches to gravity in the very early Universe.  One popular 
approach is Hawking's Euclidean path integral \cite{Hawking}, which 
describes the wave function of the Universe in terms of a ``Wick rotated'' 
gravitational path integral over Riemannian (positive definite) metrics 
$g$ on a spacetime manifold $M$, with an action 
\beq
I_E[g,\phi;M] = -\frac{1}{16\pi G}\int_M d^n x \sqrt{g}(R[g]-2\Lambda) 
              -\frac{1}{8\pi G}\int_{\partial M} d^{n-1}x \sqrt{h} K
              + I_{\rm matter}[\phi,g] .
\label{1x1}
\eeq
Here $R[g]$ is the scalar curvature, $\Lambda$ is the cosmological constant, 
$h$ is the induced metric on $\partial M$, and $K$ is the trace of the 
intrinsic curvature of $\partial M$, while $\phi$ represents a generic 
collection of matter fields.

A path integral ordinarily determines a transition amplitude between an 
initial and a final configuration, and to specify a unique wave function
one must select appropriate initial conditions.  The Hartle-Hawking ``no 
boundary'' proposal \cite{HartHawk} is that there should be {\it no\/} 
initial geometry---the path integral should be evaluated for compact 
manifolds $M$ with only a single, connected boundary component $\Sigma$. 
If we specify a metric $h$ and a set of matter fields $\phi|_\Sigma$ on 
$\Sigma$, the path integral
\beq
\Psi[h,\phi|_\Sigma;M] = \int [dg][d\phi]\, \exp\left\{-I_E[g,\phi;M]\right\}
\label{1x2}
\eeq
can be interpreted as a wave function, giving an amplitude for the universe 
with spatial topology $\Sigma$ to have an ``initial'' spatial geometry $h$ 
and matter configuration $\phi|_\Sigma$.  There are plausible, although
not conclusive, arguments that the spacetime $M$ in the path integral 
should be orientable to allow the existence of ordinary quantum field 
theory \cite{Gibbons1,Friedman}; we shall generally limit our attention to 
such manifolds in this paper.  Apart from this restriction, however, there 
seems to be no natural way to select any one particular topology in the path 
integral.  Following the general quantum mechanical prescription for
dealing with such alternative ``paths,'' the Hartle-Hawking proposal is 
therefore to sum over all manifolds $M$, subject to the condition that 
$\Sigma$ be the sole boundary component:
\beq
\Psi[h,\phi|_\Sigma;\Sigma] 
  = \sum_{M:\partial M=\Sigma}\Psi[h,\phi|_\Sigma;M] .
\label{1x2a}
\eeq

To obtain interesting physics,  a further restriction on $M$ must be 
imposed.  The integration in \rref{1x2} is over Riemannian metrics, and it 
is necessary to ``analytically continue'' to obtain the observed Lorentzian 
structure of spacetime.  This will be possible if the Riemannian metrics 
in the path integral can be joined to Lorentzian metrics to the future of 
$\Sigma$.  Gibbons and Hartle \cite{GibHart} have shown that a finite 
action continuation across $\Sigma$ exists only if the extrinsic curvature 
$K_{ij}$ of $\Sigma$ vanishes, that is, if $\Sigma$ is totally geodesic.
One should thus limit the sum to ``real tunneling geometries,'' manifolds 
$M$ and metrics $g$ such that $\partial M = \Sigma$ is totally geodesic.  

Strictly speaking, this condition does not make sense in a quantum 
theory: the boundary metric $h_{ij}$ and the extrinsic curvature $K_{ij}$ 
are canonically conjugate variables, and cannot be specified simultaneously.  
Semiclassically, this problem manifests itself in the fact that the restriction 
$K_{ij}=0$ often determines $h_{ij}$ uniquely or almost uniquely.  Real 
tunneling geometries may thus give saddle point contributions to the wave 
function $\Psi[h,\phi|_\Sigma;M]$ at only a few values of $h$; to determine 
$\Psi$ at other values, one must consider complex metrics \cite{Halliwell}, 
with consequent ambiguities in the choice of integration contour.  
Fortunately, though, it has been shown in Ref.\ \cite{Carlip_HH} that 
the restriction to real tunneling geometries still reproduces the locations 
of the extrema of the full wave function.  In particular, in the semiclassical
approximation considered below, it is not unreasonable to expect that the 
restriction to real Riemannian metrics with totally geodesic boundaries 
should correctly approximate the peaks of the wave function.

Let us now restrict ourselves to four-dimensional spacetimes with no matter
and a negative cosmological constant.  (The generalization to $n>4$ is 
straightforward, but, as we shall explain below, the generalization to
positive $\Lambda$ is not.)  The extrema of the path integral (\ref{1x2})
are then Einstein metrics $\bar g$, with actions
\beq
\bar I_E(M) = -\frac{\Lambda}{8\pi G}{\mathit{Vol}}_{\bar g}(M) 
\label{1x3}
\eeq
where ${\mathit{Vol}}_{\bar g}(M)$ is the volume of $M$ evaluated with
respect to the metric $\bar g$.  In the saddle point approximation,
the Hartle-Hawking wave function is thus
\beq
\Psi[h;\Sigma] \approx \sum_{M:\partial M=\Sigma}\Delta_M \exp
  \left\{-\frac{|\Lambda|}{8\pi G}{\mathit{Vol}}_{\bar g}(M)\right\}
\label{1x4}
\eeq
where the prefactors $\Delta_M$ are combinations of determinants coming 
from gauge-fixing and from small fluctuations around the extrema \cite{Mazur}. 

It is apparent from (\ref{1x4}) that contributions from ``large'' manifolds
are exponentially suppressed, and it is natural to guess that the wave 
function is dominated by the smallest-volume Einstein manifold with the 
specified boundary data.  In three spacetime dimensions, this was shown 
to be wrong in Refs.\ \cite{Carlip1,Carlip2}: although the contributions
of high-volume manifolds are exponentially suppressed, the number of
such manifolds can be large enough to outweigh this suppression, and 
the sum over topologies can drastically change the behavior of the wave 
function.  

In the more interesting case of four spacetime dimensions, a similar kind
of topological dominance is known for one special case \cite{RatTsch1},
and the partition function is also known to be dominated by complicated, 
high-volume manifolds \cite{Carlip3,Carlip4}.  The purpose of this paper is 
to show that this behavior is much more general: for a wide class of manifolds, 
the sum over topologies produces sharp peaks in the Hartle-Hawking wave
function that could not have been guessed by looking at any single
contribution.  Because of limits to our present understanding of
the space of Einstein metrics, a complete, systematic understanding of 
this phenomenon is still lacking, but ultimately it may be possible to use 
this sort of analysis to make testable predictions about the geometry
and topology of the Universe.

\section{Einstein Dehn Filling \label{edf}}

Our conclusions are based on an interesting new development in 
differential geometry, the discovery that one can use the method 
of Dehn filling on a cusped hyperbolic manifold to produce a large
class of topologically distinct but geometrically close Einstein 
manifolds \cite{Anderson}.  In this section, we shall briefly summarize 
this technique.

Let $N$ be a complete, hyperbolic (that is, constant negative curvature)
four-manifold of finite volume, with metric $g_N$.  Such a manifold 
need not be compact: it may instead contain cusp ends, regions 
isometric to ${\mathbb R}^+\times T^3$ with the AdS-like metric
\beq
ds^2 = \frac{3}{\Lambda}
  \left(dt^2 + e^{-2t}d{\mathbf x}\cdot d{\mathbf x}\right) ,
  \quad 0<t<\infty .
\label{2x1}
\eeq
More generally, the three-torus $T^3$ in such a cusp may be replaced 
with an arbitrary compact flat three-manifold, of which ten topological 
types exist, six of them orientable \cite{Wolf}.  We shall postpone 
consideration of such nontoral cusps until later.

If one cuts a cusp $E$ at constant $t$, the resulting manifold 
$N\backslash E$ acquires a  new boundary with the topology of a three-torus.    
``Dehn filling'' is the process of filling in this boundary by gluing on a solid 
torus $D^2\times T^2$, whose boundary is also $\partial D^2\times T^2 
\simeq T^3$.  This process is not unique: the boundary of the solid torus 
has a preferred simple closed curve, $\partial D^2$, which we may identify 
with any simple closed curve on the toral boundary of $N\backslash E$.  If 
we specify a simple closed geodesic $\sigma$ on this boundary,\footnote{Any 
simple closed curve on this $T^3$ boundary is homotopic to a geodesic.} 
however, and attach the solid torus by a diffeomorphism that identifies 
$\sigma$ and $\partial D^2$, this is sufficient to determine the topology 
of the resulting manifold.  The geodesic $\sigma$, in turn, is fixed by three 
integers ${\mathbf{n}} = (n_1,n_2,n_3)$: if one chooses three independent 
``circumferences'' $v_i$ of $T^3$ that intersect only at a single base point, 
then $\sigma$ is homotopic to a linear combination ${\mathbf{n}}\cdot
{\mathbf v}$, where ${\mathbf{n}}$ is a triple of relatively prime integers.  
Note that for ${\mathbf{n}}$ large, the length of the geodesic $\sigma$ 
becomes large.

If $N$ has only a single cusp, this process produces an infinite number of 
topologically distinct compact manifolds $N_{\mathbf{n}}$, labeled by 
triples of relatively prime integers ${\mathbf{n}}$.  If $N$ has more 
than one cusp, one can perform the same procedure on each, obtaining
manifolds $N_{\{\mathbf{n}\}}$ labeled by a collection $\{{\mathbf{n}}\}$
of such triples.  So far, the construction is purely topological.  The main 
result of Ref.\ \cite{Anderson}  is that for suitable $\{{\mathbf{n}}\}$, 
the filled manifold admits an Einstein metric $g_{\{\mathbf{n}\}}$, and 
that this metric is geometrically close to the original metric $g_N$.

Suppose $N$ has $k$ cusps $\{E_i\}$.  For a fixed Dehn
filling, the construction of the metric $g_{\{\mathbf{n}\}}$ starts with 
$k$ copies of the toral AdS black hole metric \cite{Brill} on the solid torus 
$D^2\times T^2$, each appropriately ``twisted'' to match the desired 
filling.  These metrics can be attached continuously to the hyperbolic 
metric on $N\backslash E_i$ at each cusp.  The result is not quite smooth 
at the ``seams,''  but it is shown in Ref.\ \cite{Anderson} that for suitable 
$\{{\mathbf{n}}\}$, it can be smoothed to give an Einstein metric.  

More precisely, the topological Dehn filling is determined by $k$ geodesics 
$\{\sigma_1,\dots,\sigma_k\}$, or equivalently, as described above, 
$k$ triples of relatively prime integers $\{{\bf n}_1,\dots,{\bf n}_k\}$.  
Given such a filling, let $R_{\mathit{min}}$ be the length of the shortest 
of the geodesics $\sigma_i$, in units $\Lambda=3$, and let $R_{\mathit{max}}$ 
be the length of the longest.  Then the filled manifold $N_{\{\mathbf{n}\}}$
admits an Einstein metric provided that $R_{\mathit{min}}$ is sufficiently
large and
\beq
R_{\mathit{max}} \le e^{c_0R_{\mathit{min}}{}^3} ,
\label{r1}
\eeq
where $c_0$ is a fixed constant.  Note that all of the cusps must be
filled---heuristically, an unfilled cusp is equivalent to a geodesic with
$R_{\mathit{max}}=\infty$, and (\ref{r1}) has no solutions.  On the other 
hand, by varying all of the fillings, one may easily obtain infinitely many 
manifolds satisfying (\ref{r1}), and thus infinitely many compact manifolds 
with Einstein metrics.

Ref.\ \cite{Anderson} does not construct the Einstein metric 
$g_{\{\mathbf{n}\}}$ explicitly, but proves that it is  close to the original 
metric $g_N$, in several related senses.   First observe that $N$ embeds 
as a domain in $N_{\{\mathbf{n}\}}$; in fact,  $N_{\{\mathbf{n}\}}$ with 
the core torus $\{0\}\times T^2 $ removed from each filled cusp is 
diffeomorphic to $N$.  As $\{{\bf n}\}$ grows, each core torus $\{0\}
\times T^2 $ becomes small in the metric $g_{\{\mathbf{n}\}}$, in that 
its diameter and area go to zero as $\{{\bf n}\} \rightarrow \infty$.  
In addition, the core tori recede progressively further down the cusps, 
so that the metrics $(N_{\{\mathbf{n}\}},g_{\{\mathbf{n}\}})$ converge 
smoothly to the original metric $g_{N}$ on any fixed compact set in 
$N$.  One consequence of this behavior is that the volumes converge: 
there are constants $\delta_{\{\mathbf{n}\}}>0$ such that
\beq
\mathit{Vol}_{g_{\{\mathbf{n}\}}}(N_{\{\mathbf{n}\}}) = \mathit{Vol}_{g_N}(N) -
  \delta_{\{\mathbf{n}\}} \quad \hbox{with $\delta_{\{\mathbf{n}\}}\rightarrow0$  
  as $\{{\mathbf{n}}\}\rightarrow\infty$} .
\label{2x2}
\eeq
(We refer the reader to chapter 10 of Ref.\ \cite{Petersen} for a general 
discussion of norms used to define closeness in Riemannian geometry.)

As noted above, a cusp of a hyperbolic four-manifold need not be toral, 
but can have the topology ${\mathbb R}^+\times F$, where $F$ is any 
flat three-manifold.  Any such $F$ can be obtained as a quotient of
${\mathbb R}^3$ by a group of isometries.  In Ref.\ \cite{Anderson} 
it is shown that the techniques described here for toral ends can be 
applied to four of the ten possible flat geometries of a cusp end, the 
geometries $G_1$, $G_2$, $B_1$, and $B_2$ of Ref.\ \cite{Wolf}, or
equivalently $A$, $B$, $G$, and $H$ of Ref.\ \cite{RatTsch0}.  Of these,
only cusps of type $A$ and $B$ are orientable.

\section{The Partition Function \label{pf}}

One immediate consequence of this construction is that the semiclassical
gravitational partition function $Z$---the path integral (\ref{1x1}) for
{\it closed\/} four-manifolds---diverges when $\Lambda<0$.  Indeed, 
given any hyperbolic ``seed manifold'' $N$ with the appropriate cusp 
types, Einstein Dehn filling produces an infinite number of compact 
Einstein manifolds, all with volumes less than ${\mathit{Vol}}_{g_N}(N)$.  
Since the semiclassical contribution of each such manifold goes as
\beq
\exp \left\{-\frac{|\Lambda|}{8\pi G}
  \mathit{Vol}_{g_{\{{\mathbf{n}}\}}}(N_{\{\mathbf{n}\}}) \right\} ,
\label{2.5x1}
\eeq
just as in (\ref{1x4}), the sum over the $N_{\{\mathbf{n}\}}$ is infinite, at 
least at this order of approximation.  The possibility remains, of course, 
that higher-order corrections will strongly suppress all but a few of the 
contributions to the sum.  There seems to be no reason to expect that, 
however, particularly since all of the Einstein metrics 
$(N_{\{\mathbf{n}\}},g_{\{\mathbf{n}\}})$ are strictly stable---that is, the
second variation of the action has no zero or negative eigenmodes, which
could otherwise lead to bad behavior or introduce phases in the prefactors 
$\Delta$ in (\ref{1x4}).

The fact that the partition function diverges is not new.  It was shown 
in Refs.\ \cite{Carlip3,Carlip4} that even the contributions of strictly 
hyperbolic manifolds lead to a divergent sum, since the number of 
manifolds with volume greater than $V$ grows faster than exponentially 
with $V$ \cite{Burger}.  That sum, however, is similar to a divergent
sum occurring in string theory \cite{Gross}, and one might hope that 
a finite resummation similar to that coming from matrix models in
string theory \cite{matrix} might be possible.  The divergence found
here, on the other hand, has no obvious cure, although we shall
discuss a few possible loopholes in section \ref{candc}.

The physical meaning of the partition function, however, is not entirely
clear.  If one restricts oneself to manifolds of the form $S^1\times\Sigma$,
or more generally to circle bundles over a three-manifold, one can
interpret the circumference of the $S^1$ factor as a local inverse temperature
and $Z$ as a thermal partition function.  The manifolds we consider
here are not of that form, however, and it is not obvious that a divergent 
partition function is a terrible thing.  It is therefore important to look 
at quantities with clearer physical interpretations, such as the Hartle-%
Hawking wave function (\ref{1x2a}).

\section{Totally Geodesic Boundaries \label{tgb}}

The discussion in Ref.\ \cite{Anderson} was restricted to manifolds with
no boundaries apart from cusps.  To understand the Hartle-Hawking
wave function, we must instead consider manifolds with a totally geodesic 
boundary $\Sigma$.  Fortunately, the generalization is fairly straightforward.

Let $M$ be a finite-volume hyperbolic manifold with cusps that has a connected 
totally geodesic boundary $\Sigma$, with induced metric $h_M$ on $\Sigma$.  
The double $N=2M$ is formed by attaching two copies of $M$ along $\Sigma$.  
This double can be characterized by the existence of a reflection isometry
$\theta$ that permutes the two copies of $M$, with a fixed point set $\Sigma$, 
the ``mirror'' of the reflection. The existence of such an isometry with a 
two-sided, separating fixed point set ensures that $2M$ can be cut along 
this mirror to obtain the original manifold $M$ with totally geodesic boundary 
$\Sigma$.  If $M$ is orientable, $\theta$ will be orientation-reversing.

The double $2M$ meets the conditions of section \ref{edf}, and we can 
perform an identical Dehn filling on each pair of cusps related by the
reflection symmetry.  The smoothing procedure of Ref.\ \cite{Anderson}
respects this symmetry, and thus produces a manifold $2M_{\{\mathbf{n}\}}$
that itself has a reflection symmetry.  Cutting $2M_{\{\mathbf{n}\}}$ 
along the fixed point set of this symmetry, we obtain a manifold 
$M_{\{\mathbf{n}\}}$ with an Einstein metric $g_{\{\mathbf{n}\}}$ and 
a totally geodesic boundary $\Sigma_{\{\mathbf{n}\}}$.  Since the Dehn 
filling affects the topology only at the cusp ends, the boundary 
$\Sigma_{\{\mathbf{n}\}}$ is diffeomorphic to $\Sigma$.  Furthermore, 
the convergence conditions described in section \ref{edf} guarantee that 
the induced boundary metrics $h_{\{\mathbf{n}\}}$ converge to $h_M$, 
and that the volume $\mathit{Vol}(M_{\{\mathbf{n}\}})$ converges to 
$\mathit{Vol}(M)$ as in (\ref{2x2}).

More care is needed if the double $2M$ has cusps that intersect the mirror 
hypersurface $\Sigma$.  Let $\{E\}$ denote the cusps that intersect $\Sigma$, 
and let $\{{\widehat E}\}$ denote the cusps that do not.  We can then look
for a fixed Einstein Dehn filling of the $\{E\}$ that respects the mirror symmetry 
and proceed as above.  If such a symmetric filling exists,\footnote{See 
Appendix B for a detailed description of some examples.}  it will change 
the topology of $\Sigma$; for a fixed symmetric filling of the $\{E\}$ labeled 
by integers $\{\mathbf{m}\}$, we will obtain a new, compact boundary 
$\Sigma^{\{\mathbf{m}\}}$.  

By now varying the fillings of the remaining cusps $\{{\widehat E}\}$, we 
can construct a large number of Einstein four-manifolds with induced 
boundary metrics that are close to a fixed $h_{M^{\{\mathbf{m}\}}}$, and 
with volumes close to a fixed $\mathit{Vol}(M^{\{\mathbf{m}\}})$.  Because
of the requirement (\ref{r1}), the number of such manifolds is no longer
infinite: the fixed filling of the $\{E\}$ determines a minimum length
$R_{\mathit{min}}$, and (\ref{r1}) then restricts the fillings of the remaining
cusps $\{{\widehat E}\}$.  The number of allowed fillings grows very rapidly
with $R_{\mathit{min}}$, however; for a double $2M$ with $2k$ cusps that
do not intersect the mirror hypersurface, the number of allowed fillings
goes as $\exp\{3kc_0R_{\mathit{min}}{}^3\}$.  Since $R_{\mathit{min}}$ 
is typically very large, this number quickly becomes enormous.

The next question we must address is the existence of ``seed manifolds''
$M$ of the type described here.  Unfortunately, at least for now we cannot 
completely classify such manifolds.  There are, however, two constructions 
that show that such manifolds exist.  The first and most explicit of these 
is based on Ref.\ \cite{RatTsch0}, in which 1171 complete hyperbolic 
four-manifolds of minimum volume are constructed by gluing together 
the sides of a regular ideal 24-cell.  The manifolds of this class all contain 
totally geodesic hypersurfaces, and as described in \cite{RatTsch2}, cutting 
along such a hypersurface will sometimes result in a complete hyperbolic 
manifold with a single totally geodesic boundary.  For our construction to 
work, we must further check that the resulting manifold has the proper 
cusp types to allow the Einstein Dehn filling of Ref.\ \cite{Anderson}.

By examining the manifolds of Ref.\ \cite{RatTsch0}, we have found 
a large number of such manifolds.  Most of these, however, have cusps
of type $B$ that intersect the mirror $\Sigma$.  As we explain in Appendix 
B, such cusps do not admit Einstein Dehn fillings that respect the reflection 
symmetry, and hence cannot be filled to provide ``seed manifolds''  of the 
type we need. There are, however, five examples---described in detail in 
Appendix A---for which only cusps of type $A$ intersect the mirror $\Sigma$, 
giving us precisely the structure we need.

Consider, for example, manifold $N_{24}$ of Appendix A.  This manifold 
is the orientable double cover of manifold 24 of \cite{RatTsch0}, and has 
Euler characteristic two and a volume of $24\pi^2/\Lambda^2$.  It has 
seven cusps, all of type $A$, three of which intersect the mirror.  The mirror 
$\Sigma$ in $N_{24}$, which will become the ``spatial'' boundary 
in the Hartle-Hawking wave function, is a hyperbolic three-manifold
homeomorphic to a double cover of the complement of the four-component 
link  $8^4_2$.  It has six cusps, and is therefore not compact.  We 
can, however, make $\Sigma$ compact by performing an Einstein Dehn
filling of the three cusps of $N_{24}$ that intersect $\Sigma$.  As we discuss in 
Appendix B, it is possible to do this in a manner that respects the reflection 
symmetry of $N_{24}$, ensuring that the filled manifold is still a double.  In
fact, such a filling induces an ordinary (topological) Dehn filling of 
$\Sigma$.  Conversely, any sufficiently large Dehn filling of $\Sigma$ that 
has the symmetries described in Appendix B can be lifted to an Einstein 
Dehn filling of $N_{24}$.  By choosing a large enough fixed filling of the
cusps of $\Sigma$, one can make $R_{\mathit{min}}$ arbitrarily large, 
and can thus find arbitrarily many fillings of the remaining cusps that 
satisfy (\ref{r1}).

The remaining manifolds described in Appendix A are different in detail, 
but qualitatively similar.  Each has Euler characteristic two and a volume 
of $24\pi^2/\Lambda^2$.  Each is a double, with a mirror $\Sigma$ 
that has cusps; in fact, the topology of this mirror is identical in all
five examples.  In each case, the cusps of $\Sigma$ can be filled by  
Einstein Dehn fillings that respect the reflection symmetry.  By choosing 
different Dehn fillings of the cusps that intersect the mirror, as described 
in Appendix B, we can thus obtain an infinite collection of boundary 
topologies for which our construction applies.

Another set of potential ``seed manifolds'' are due to Long 
and Reid \cite{LongReid1}, who give algebraic descriptions of a variety 
of hyperbolic four-manifolds with totally geodesic boundaries.  The 
manifolds discussed in this reference are compact, but the results may 
be extended to manifolds with cusps \cite{Reid}.  Further investigation 
of these examples would be of interest.

Long and Reid have also found a large set of three-manifolds for which
our construction {\it cannot\/} work.  In Ref.\ \cite{LongReid2}, they
show that a three-manifold $\Sigma$ can occur as a totally geodesic
boundary of an orientable hyperbolic four-manifold only if the eta invariant 
$\eta(\Sigma)$ is an integer.  As it is written, theorem 1.1 of that paper 
applies only to boundaries of compact orientable four-manifolds.  But it 
follows from the proof of theorem 1.3 that for an orientable hyperbolic 
four-manifold $M$ with totally geodesic boundary $\Sigma$ and cusps 
with topologies ${\mathbb R}^+\times F_i$,
\beq
\eta(\Sigma) + \sum_i\eta(F_i) \in \mathbb{Z} .
\label{3x1}
\eeq
Long and Reid further show that $\eta(F_i)$ is an integer for cusps of type
$A$ and $B$, the two for which Einstein Dehn fillings are possible.   Hence 
for our construction to apply, we must require that $\eta(\Sigma)\in\mathbb{Z}$.  
This is a strong restriction: of some $11,000$ small-volume hyperbolic 
three-manifolds in the SnapPea census \cite{Snap}, for example, only 41 have 
integral eta invariants, so only these 41 are even candidates for our construction.
(It is not clear that this result generalizes in any easy way if one includes
nonorientable four-manifolds, but as argued in section \ref{intro},  
there are good reasons to expect only orientable four-manifolds to
contribute to the Hartle-Hawking wave function.)
 
\section{Peaks in the Hartle-Hawking Wave Function \label{hhp}}

We now turn to the implications of these results for the
Hartle-Hawking wave function.  Consider a universe described, at least 
in the low energy limit, by general relativity with a negative cosmological 
constant, as required, for example, by supergravity.  Let $\Sigma$ be a 
three-manifold of the sort discussed in the preceding section---that is, a 
manifold that can occur as a totally geodesic boundary of a hyperbolic 
manifold $M$ with cusps of type $A$ or $B$ (or $G$ or $H$ if we permit
nonorientable manifolds).  Denote by $h_M$ the metric induced on 
$\Sigma$ by the hyperbolic metric $g_M$ of $M$.

The manifold $M$ will occur as a saddle point in the Hartle-Hawking path
integral (\ref{1x2}), giving a contribution, to lowest order, of
\beq
\exp\left\{-\frac{|\Lambda|}{8\pi G}{\mathit{Vol}}_{g_M}(M)\right\} 
\label{4x1}
\eeq
at the boundary value $h_M$.  But it follows from the discussion above 
that there are a very large number of Einstein manifolds $M_{\{\mathbf{n}\}}$, 
also with boundary $\Sigma$, whose metrics have boundary values 
$h_{\{\mathbf{n}\}}$ that lie close $h_M$, in the sense of ``closeness''  
discussed in section \ref{edf}.  By (\ref{2x2}), the volumes of these filled 
manifolds are all less than ${\mathit{Vol}}_{g_M}(M)$, so at this order 
their contribution to the wave function is actually greater than that of $M$.  

A large number of manifolds thus give nearly identical contributions to the
sum (\ref{1x4}) within a small neighborhood of the boundary metric $h_M$.  
Since the standard Hartle-Hawking prescription (\ref{1x2a}) tells us to simply 
add such contributions, the Hartle-Hawking wave function for $\Sigma$, at 
this order of approximation, is therefore sharply peaked around $h_M$.  Note 
that such a peak occurs for any value of $\Lambda<0$ and for any finite 
volume of the ``seed manifold'' $M$, and will appear for every spatial topology
$\Sigma$ that occurs as a boundary of an appropriate four-manifold.

If we could find a hyperbolic ``seed manifold'' whose double had no cusps
on the mirror $\Sigma$, this peak would, in fact, be infinite.  As discussed 
in the preceding section, however, the ``seed manifolds'' that we know 
how to construct explicitly all have boundaries $\Sigma$ with cusps.  To
apply the construction of \cite{Anderson}, we must fill these cusps, as 
discussed above and in Appendix B.  If a fixed filling of these cusps respects 
the reflection symmetry, it gives a new ``seed manifold'' $M^{\{\mathbf{m}\}}$ 
with boundary $\Sigma^{\{\mathbf{m}\}}$ related to $\Sigma$ by a topological
Dehn filling.  In the examples in Appendix A, such a partially filled four-manifold 
will still have a set of cusps that do not intersect $\Sigma$, and summing over 
the allowed fillings of these remaining cusps will  give us a sharp peak in the 
Hartle-Hawking wave function on $\Sigma^{\{\mathbf{m}\}}$.  

Note that while the $M^{\{\mathbf{m}\}}$ are not exactly hyperbolic, they
are ``nearly'' hyperbolic in the sense of section \ref{edf}.  In particular, for 
large $\{\mathbf{m}\}$ the induced metric on $\Sigma^{\{\mathbf{m}\}}$ 
will be very close to the hyperbolic metric on $\Sigma$ except in regions
very far down the filled cusps.  Moreover, the more nearly hyperbolic 
$M^{\{\mathbf{m}\}}$, the larger the geodesic length $R_{\mathit{min}}$
in (\ref{r1}) will be, and the more Einstein Dehn fillings will be allowed.
``Nearly hyperbolic'' boundaries thus have very highly peaked wave 
functions.

Let us now point out an intriguing feature of this construction.  The 
peaks that we have found are centered around metrics $h_M$ (or 
$h_{M^{\{\mathbf{m}\}}}$) obtained from hyperbolic or very nearly 
hyperbolic ``seed manifolds.''  Since the boundary $\Sigma$ is always
a hypersurface of vanishing extrinsic curvature, such an $h_M$ is itself a 
hyperbolic or very nearly hyperbolic metric on $\Sigma$.  This means 
that the peaks occur at or very near to locally homogeneous geometries
of $\Sigma$.  Of course, we do not know that these are the {\it only\/} 
peaks of the Hartle-Hawking wave function.  Nevertheless, the appearance 
of a new, noninflationary mechanism for explaining spatial homogeneity 
of the early universe should be of some interest.

\section{Conclusions and Caveats \label{candc}}

The calculations we have described so far are, of course, only lowest order 
approximations to the full path integral (\ref{1x2})--(\ref{1x2a}).  To the next 
order, we should take into account the prefactors $\Delta_{M_{\{\mathbf{n}\}}}$ 
of (\ref{1x4}).  The convergence of metrics is not, unfortunately, enough to 
guarantee the convergence of these determinants, which depend nonlocally 
on the metric throughout $M_{\{\mathbf{n}\}}$.  In three spacetime dimensions, 
the prefactors combine to form a topological invariant, the Ray-Singer torsion 
\cite{Witten}, which does not converge under a similar Dehn filling procedure 
\cite{Carlip1}.  In that case, however, the prefactors densely fill a finite 
nonnegative interval, a behavior sufficient to show that the first-order infinite 
peak survives.  

For four spacetime dimensions, no corresponding result is known, and it 
remains possible, although unlikely, that the prefactors become systematically 
small enough for large $\{\mathbf{n}\}$ to suppress the peaks we have found.  As 
noted in section \ref{pf}, the strict stability of our filled manifolds implies the
absence of negative eigenvalues that could otherwise lead to nontrivial phases 
\cite{Hawking}, eliminating one possible source of suppression.  We can actually 
say a bit more about one-loop corrections.  In an effective action formalism 
\cite{DeWitt}, such corrections lead to new terms of the form
\beq
\int_{M_{\{\mathbf{n}\}}} d^4x \sqrt{g}C_{abcd}C^{abcd}
\label{5x1}
\eeq
in the effective action, where $C_{abcd}$ is the Weyl tensor.  But such terms
converge to zero for large $\{\mathbf{n}\}$ \cite{Anderson}, and thus do not 
affect our conclusions.  Indeed, integrals of higher order polynomials in the 
Weyl tensor, which can appear at higher orders of perturbation theory, likewise
converge to zero.  Quantum corrections can also induce terms in the action
(\ref{1x1}) proportional to the Euler characteristic $\chi$ and the Hirzebruch
signature $\tau$ \cite{Hawking}.  The Euler characteristic is left invariant
by Dehn filling, however, and the signature can be written in terms of
integrals of the Weyl curvature that again go to zero for large fillings.  Hence
neither of these topological terms will affect our conclusions.  While this
does not guarantee the absence of higher-order nonlocal effects that could 
still suppress the peaks we have found, it makes our results considerably
more robust.

One more loophole remains.  We have assumed that the sum (\ref{1x2})
over topologies should include every manifold with a boundary $\Sigma$,
and that all occur with equal weights.  While this is a plausible assumption, 
it may not be correct.  Sorkin et al.\ \cite{Dowker,Borde} have suggested 
that considerations of causal structure may restrict the topologies 
occurring in the sum over histories, and that such a restriction may 
eliminate problems of infinite particle production \cite{Anderson2} in 
topology-changing amplitudes.  The restriction to ``causally continuous'' 
spacetimes requires, in particular, that the first Betti number $b_1$ of 
the spacetime relative to the boundary vanish.  This condition would 
eliminate many of the examples considered here; we are currently investigating 
the question of whether an infinite number would remain.  Observe 
also that the manifolds in our sum over topologies with large Dehn fillings 
differ significantly from each other only in extremely small regions ``far
down the cusps.''  If the correct quantum theory of gravity contains a 
cut-off in volume or diameter, say at the Planck scale, then perhaps these 
manifolds should be considered equivalent and not counted separately.  
Until we have a much better established quantum theory of gravity, these 
issues are likely to remain unresolved.

Assuming that the manifolds we have considered here really do appear 
in the sum over topologies, an analysis of this sort may eventually have
predictive value, telling us what spatial geometries of the Universe are 
most probable.  For now, though, several sizable obstacles remain.  

First, of course, we do not have any particular reason to believe that the
cosmological constant is negative.  A negative $\Lambda$ in the very
early Universe is not excluded by observation---the apparently positive 
cosmological constant now may be a consequence of later dynamics---but
neither is it supported.  The analysis of this paper relies on particular
properties of manifolds with negative curvature, and does not extend
to those with positive curvature.  In three spacetime dimensions, the case
of positive $\Lambda$ can be analyzed, and is qualitatively very different
\cite{Carlip2}.  Ideally, one would like to do the same in four spacetime
dimensions, but to the best of our knowledge the appropriate mathematical 
tools are not yet available.

Second, we do not yet have good control over which manifolds can appear 
in the constructions of this paper.  In particular, while section \ref{tgb}
provides examples of three-manifolds $\Sigma$ that can occur as
boundaries in an Einstein Dehn filling, we have no reason to expect these 
examples to constitute a complete list.  We clearly need a much better grasp of 
which three-manifolds can occur as totally geodesic boundaries of complete
hyperbolic four-manifolds.

Third, while the method of Einstein Dehn fillings gives some peaks of
the Hartle-Hawking wave function, it may not give all of the peaks.
In general, the Hartle-Hawking prescription tells us to start with
extrema of the action (\ref{1x1}).  But while such extrema are always
Einstein manifolds, they need not be hyperbolic or ``nearly'' 
hyperbolic manifolds of the type we have discussed in this paper.  For 
example, it is tempting to conclude from the results of Long and Reid 
described at the end of section \ref{tgb} that a three-manifold $\Sigma$ 
with $\eta(\Sigma)\not\in {\mathbb Z}$ cannot have a peaked wave 
function.  But this may not be true: while such a manifold cannot occur 
as a totally geodesic boundary of a hyperbolic spacetime, we do not know 
whether it can occur as a totally geodesic boundary of a nonhyperbolic 
Einstein spacetime, or, if so, how many four-manifold topologies can
contribute.  Ultimately, we need a much more thorough understanding 
of the space of Einstein metrics.

Despite these limitations, though, we believe that this work has some
importance.  We have demonstrated that the sum over topologies can
have a clear and dramatic effect on the Hartle-Hawking wave function
in the physically realistic case of four spacetime dimensions, and that
it leads to at least some sharp peaks at spatially homogeneous geometries.  
And in the unlikely but still possible event that observations of the cosmic 
microwave background \cite{Cornish} or other of cosmic structure
\cite{Roukema} demonstrate that our Universe has one of the spatial 
topologies described in section \ref{tgb}, we have a genuine prediction 
of a highly probable initial geometry.

\newpage
\begin{flushleft}
\large\bf Acknowledgments
\end{flushleft}

We would like to thank Gary Gibbons and Alan Reid for valuable help.  
This work was supported in part by U.S.\ Department of Energy grant 
DE-FG03-91ER40674 and NSF grant DMS 0305865.

\appendix
\section{Seed Manifolds}

In this appendix we describe the ``seed manifolds'' of section \ref{tgb}.
We start with the complete minimum-volume hyperbolic manifolds of 
Ref.\ \cite{RatTsch0},  which were obtained by gluing together the sides 
of a single regular ideal 24-cell.  As described in \cite{RatTsch2}, each of 
these manifolds contains a collection of totally geodesic hypersurfaces,
or ``cross sections,'' inherited from the hyperplanes of symmetry of
the 24-cell.  If $N$ is a nonorientable manifold chosen from this list
with a cross section $S$, one can cut $N$ along $S$ to obtain a manifold
$M$ with boundary $\Sigma$.  Ref.\ \cite{RatTsch2} gives the criteria for 
$M$ to be a ``seed manifold'' of the type we need: $S$ must be orientable, 
one-sided, and have an orientable complement.  For the Einstein Dehn 
filling of Ref.\ \cite{Anderson} to apply, we must also demand that the 
cusps of the double $2M$ all be of type $A$ or $B$.  Moreover, as explained 
in Appendix B, any cusps of $2M$ that intersect the mirror $\Sigma$
must be of type $A$.

In table \ref{tab1} we give five examples that meet these criteria, together 
with the homology and cusp link types of their doubles.  The column headed 
$N$ gives the number from \cite{RatTsch0} of the nonorientable four-manifold 
$N$ that we take as a starting point.  The column headed by $\mathit{SP}$ 
gives the side-pairing code for gluing the ideal 24-cell to obtain $N$, as 
explained in \cite{RatTsch0}.  The pairings are taken so that the coordinate 
hyperplane cross section $x_{4}=0$ is a one-sided, nonseparating, totally 
geodesic hypersurface whose complement is orientable. Cutting open $N$ 
along this cross section yields an orientable hyperbolic four-manifold 
$M$ with a connected totally geodesic boundary $\Sigma$. The boundary 
$\Sigma$ double covers the cross section of the starting four-manifold.  
The reflection in the $x_{4}=0$ cross section is a symmetry of $N$, and 
induces a symmetry of $M$ that restricts to the covering transformation of 
$\Sigma$. This implies that the double of $M$ along $\Sigma$ is the orientable 
double cover of $N$.

\begin{table}[htp]
\begin{center}
\begin{tabular}{rccccl}
$N$&$\mathit{SP}$&$H_{1}$&$H_{2}$&$H_{3}$&$\mathit{LT}$\\
{\tt 24}&{\tt 9FC129}&${\mathbb Z}^{7}\oplus{\mathbb Z}_{2}$
&${\mathbb Z}^{14}$&${\mathbb Z}^{6}$&{\tt AAAAAAA}\\
{\tt 55}&{\tt 9FC12A}&${\mathbb Z}^{5}\oplus{\mathbb Z}_{2}^{3}$
&${\mathbb Z}^{12}$&${\mathbb Z}^{6}$&{\tt AAAAABB}\\
{\tt 237}&{\tt 9FC179}&${\mathbb Z}^{3}\oplus{\mathbb Z}_{2}^{5}$
&${\mathbb Z}^{10}$&${\mathbb Z}^{6}$&{\tt AAABBBB}\\
{\tt 1091}&{\tt 9FC124}&${\mathbb Z}^{9}$
&${\mathbb Z}^{18}$&${\mathbb Z}^{8}$&{\tt AAAAAAAAA}\\
{\tt 1113}&{\tt 9FC127}&${\mathbb Z}^{3}\oplus{\mathbb Z}_{2}^{6}$
&${\mathbb Z}^{12}$&${\mathbb Z}^{8}$&{\tt AAABBBBBB}\\
\end{tabular}
\end{center}
\caption{Five double 24-cell manifolds.\label{tab1}}
\end{table}

In each of these examples, the coordinate hyperplane cross section is the 
hyperbolic three-manifold $M_{10}^{3}$ described in \cite{RatTsch0} with 
side-pairing code $174$ for the gluing of the sides of an ideal rhombic 
dodecahedron.  The ideal rhombic dodecahedron is the coordinate hyperplane 
cross section of the ideal 24-cell fundamental domain.  The pairings of 
sides of the ideal 24-cell for our starting four-manifold $N$ that give rise 
to the side-pairing of the rhombic dodecahedron cross section all reflect 
in the direction  perpendicular to the cross section, thus joining the top 
and bottom sides of the cross section in the 24-cell.  This implies that the 
boundary $\Sigma$ of $M$ can be obtained by gluing two copies of the ideal 
rhombic dodecahedron fundamental domain of $M_{10}^{3}$ so that each side 
glues to a side of the other dodecahedron by the gluing that applies for 
$M_{10}^{3}$.  See Figure 5 in \cite{RatTsch0} for an illustration of the 
ideal rhombic dodecahedron fundamental domain of $M_{10}^{3}$. 

The mirror $\Sigma$ has six cusps, each of whose link is a torus. The first 
and second homology groups of $\Sigma$ are $H_{1}(\Sigma) ={\mathbb Z}^{6}$ 
and $H_{2}(\Sigma)={\mathbb Z}^{5}$.  In units such that $\Lambda=-2$, 
the volume of $\Sigma$ is $16L(2) = 14.6554494\ldots.$, where $L(s)$ is the 
Dirichlet $L$-function  
\beq
L(s) = 1-\frac{1}{3^s}+\frac{1}{5^s}-\frac{1}{7^s}+\cdots.
\label{apa1}
\eeq

Doubling the four-manifold $M$ by reflecting across the mirror $\Sigma$ 
gives the orientable double cover of our starting nonorientable four-manifold
$N$.  For a more explicit description, we can read off a side-pairing of two 
ideal 24-cells from the side-pairing code of $N$: we use the original side-pairing,
but glue a given side to the appropriate side of the same 24-cell if the 
side-pairing code is at most 7, and to the corresponding side of the other 
24-cell if the side-pairing code is 8 or more in hexadecimal notation. 

The columns of Table \ref{tab1} headed $H_i$ list the $i$th homology groups 
of the manifolds.  The column headed $\mathit{LT}$ lists the link types
of the cusps of the manifolds. Here $A$ and $B$ represent the first two 
closed orientable flat three-manifolds in the order given by Wolf \cite{Wolf}:
$A$ is the three-torus and $B$ is the half-twisted three-torus.  See Figure 1 
of \cite{RatTsch2} for an illustration of a fundamental domain for the flat 
three-manifold $B$. 

The manifolds in the table have been selected so that only link type $A$ 
appears for the three cusps intersecting the mirror (the first three listed), 
and only $A$ and $B$ link types appear for the remaining cusps 
off the mirror.  Each cusp that intersects the mirror contains two cusps 
of the mirror.  Each link of a cusp that intersects the mirror is itself a 
double of a flat three-manifold with a totally geodesic boundary.   By 
Theorem 1 of \cite{RatTsch2}, the link of a cusp that intersects the mirror 
is the double of a right cylinder $T^2 \times [0,1]$ whose boundary components 
are the links of the cusps of the mirror that are contained in this cusp. 

The five hyperbolic four-manifolds in Table \ref{tab1} have Euler 
characteristic two, and so have volume $8\pi^2/3$ by the Gauss-Bonnet 
theorem, in units $\Lambda=-3$.  These manifolds have the smallest 
possible volume for a hyperbolic four-manifold that is a double of a 
hyperbolic four-manifold with a totally geodesic boundary. 

\section{Filling Cusps on the Mirror}

The ``seed manifolds'' of Appendix A all have cusps that intersect the 
mirror hypersurface, which therefore itself has cusps.  The construction 
of Ref.\ \cite{Anderson}, on the other hand, requires that all cusps, 
including those that intersect the mirror, be filled.  Here we describe 
the conditions for such a filling to respect the mirror symmetry.

Start with a ``seed manifold'' $N$ with a mirror $\Sigma$ fixed by a 
reflection isometry $\theta$.  Let $E\simeq{\mathbb R}^+\times F$ 
be a cusp of $N$ that intersects $\Sigma$.  Such a cusp must intersect 
$\Sigma$ symmetrically, so $\theta$ is also a reflection isometry of $F$.
The means that $F$ is itself the double of a manifold with boundary, or a
``flat gravitation instanton'' in the sense of \cite{RatTsch2}.  Such manifolds 
have been classified in \cite{RatTsch2}: if $F$ is type $A$ it must be a flat 
three-torus obtained by doubling a flat cylinder $ T^2\times[0,1]$, while if 
it is type $B$ it must be a flat closed three-manifold obtained by doubling a 
twisted $I$-bundle over a flat Klein bottle.

Suppose first that $F$ is of type $A$.  We can write $F\simeq T^2\times S^1$,
where the last $S^1$ is to be understood as the double of $[0,1]$.  The fixed 
points of the action of $\theta$ on this $S^1$ are the endpoints $\{0\}$ and 
$\{1\}$ of the interval, so $F$ intersects $\Sigma$ twice, at $T^2\times\{0\}$ 
and $T^2\times\{1\}$.  Hence a cusp of type $A$ in $N$ that intersects $\Sigma$ 
does so at two cusps of $\Sigma$.

Now, recall that a Dehn filling of $E$ requires that we identify the 
boundary of a solid torus with $F$.  Let us think of the solid torus as
$(D^2\times S^1)\times S^1$.  We can then fill the cusp $E$ in a way that 
respects the reflection symmetry by gluing $\partial D^2\times S^1$ to 
the $T^2$ factor in $F$, with $\partial D^2$ identified with some geodesic 
in $T^2$, and gluing the second $S^1$ factor in the solid torus to the $S^1$ 
in $F$ by the identity.  Such a filling is determined by two relatively prime 
integers $(m_1,m_2)$ that label the geodesic in $T^2$.  Note that this 
filling of a cusp of the four-manifold $N$ is automatically also a Dehn 
filling of the two cusps ${\mathbb R}^+\times T^2\times\{0\}$ and 
${\mathbb R}^+\times T^2\times\{1\}$ of the mirror $\Sigma$, with
surgery invariants $(m_1,m_2)$ and $(m_1,-m_2)$ that are identical
up to a sign change coming from a difference of orientation.  Conversely, 
any such symmetric filling of these two cusps of $\Sigma$ determines 
a filling of $E$.

Now suppose instead that $F$ is of type $B$.  In Ref.\ \cite{Anderson},
an Einstein Dehn filling of such a cusp was defined by first going to
a covering space, a flat manifold of type $A$, and then considering
fillings that respect the covering projection.  For a cusp that intersects
$\Sigma$, however, we must also demand that the filling respect the
reflection symmetry.  These two requirements are nearly incompatible; 
it can be shown that only two fillings are allowed, and there is no reason 
to expect these to admit Einstein metrics.  We must therefore discard 
manifolds for which cusps of type $B$ intersect the mirror.

Finally, we return to the ``seed manifold'' $N$, and now assume that 
only cusps of type $A$ intersect $\Sigma$.  We divide the cusps  of $N$ 
into two sets: $n$ cusps $\{E\}$, all of type $A$, that intersect $\Sigma$ 
and $\hat n$ cusps $\{{\widehat E}\}$, of type $A$ or $B$,  that do not 
intersect $\Sigma$.  Pick a {\it fixed\/} Einstein Dehn filling of the $\{E\}$  
that respects the reflection symmetry of $N$.  Such a filling is determined 
by $n$ pairs of relatively prime integers $\{\mathbf{m}\}$, and automatically 
fills the $2n $ cusps of $\Sigma$.

The resulting manifold $N^{\{\mathbf{m}\}}$ has a reflection isometry, 
with a mirror $\Sigma^{\{\mathbf{m}\}}$ that is now compact. Furthermore, 
$N^{\{\mathbf{m}\}}$ still has $\hat n$ cusps that do not intersect 
$\Sigma^{\{\mathbf{m}\}}$.  The allowed fillings of these remaining
cusps are restricted by (\ref{r1}), where for most fillings $R_{\mathit{min}}$
is determined by the $\{\mathbf{m}\}$.  For a ``typical'' example, 
$\{\mathbf{m}\}$ is large, $R_{\mathit{min}}\gg1$, and the number
of such fillings increases exponentially with $R_{\mathit{min}}{}^3$.
Thus, as in section \ref{hhp}, summing over Einstein Dehn fillings of these 
remaining cusps will lead to a sharp peak in the Hartle-Hawking wave 
function $\Psi[h;\Sigma^{\{\mathbf{m}\}}]$.

This peak is centered on the metric $h^{\{\mathbf{m}\}}$ induced from 
the metric of the partially filled manifold $N^{\{\mathbf{m}\}}$, which is 
not quite hyperbolic.  But for ``most'' manifolds obtained in this manner---that 
is, for large ${\{\mathbf{m}\}}$---$h^{\{\mathbf{m}\}}$ is arbitrarily close 
to a hyperbolic metric on compact sets, that is, not too far down the cusps.
In particular, the regions with large deviations from homogeneity become
arbitrarily small: for any fixed $\epsilon$, the volume of the region of
$\Sigma$ in which $h^{\{\mathbf{m}\}}$ differs from a hyperbolic metric
by more than $\epsilon$ goes to zero as $\{\mathbf{m}\}$ goes to infinity.

\end{document}